\documentclass[a4paper,12pt]{article}

\usepackage{ifpdf}

\newif\ifpdf
\ifx\pdfoutput\undefined
  \pdffalse
\else
  \pdfoutput=1
  \pdftrue
\fi

\RequirePackage{xspace} %
\RequirePackage{subfigure} %
\RequirePackage[centertags]{amsmath} %
\RequirePackage{amssymb}
\RequirePackage{wrapfig} %
\RequirePackage{calc} %
\RequirePackage{ifthen}
\RequirePackage{tabularx} %
\RequirePackage{flafter} %
\RequirePackage{fancyhdr} %

\ifpdf
  \RequirePackage[pdftex]{color}%
  \RequirePackage{colortbl}%
  \RequirePackage{array}%
  \RequirePackage[pdftex]{graphicx}

  \RequirePackage[ pdftex, plainpages = false, pdfpagelabels,
                 pdfpagelayout = useoutlines,
                 bookmarks,
                 breaklinks = true,
                 linktocpage,
                 pagebackref,                      
                 colorlinks = true,
                 linkcolor = blue,
                 urlcolor  = blue,
                 citecolor = blue,
                 anchorcolor = blue,
                 hyperindex = true,
                 hyperfigures
                 ]{hyperref}

\else
  \RequirePackage{color}
  \RequirePackage{colortbl}
   \RequirePackage{array}
  \RequirePackage[dvips]{graphicx}
  \RequirePackage{hyperref}
  \usepackage{rotating}
\fi

\usepackage{makeidx} 
\usepackage{setspace} 
\usepackage{rotating} 
\usepackage{ecltree}
\usepackage{epic}
\usepackage{supertabular}  
\usepackage{color}
\usepackage{exscale}
\usepackage{fontenc}
\usepackage{ifthen}
\usepackage{latexsym}
\usepackage{makeidx}
\usepackage{syntonly}
\usepackage{inputenc}
\usepackage{graphicx}
\usepackage{setspace}
\usepackage{caption2}
\usepackage[english]{babel}
\usepackage[square, comma,numbers,sort&compress]{natbib}
\usepackage{hypernat}
\usepackage{boxedminipage}
\usepackage{framed}
\usepackage{longtable}
\usepackage[all]{hypcap}    
\usepackage{algorithm2e}
\usepackage{algorithmic}
\usepackage{lscape}
\usepackage{pdflscape}

\newcommand{\note}[1]{\marginpar[left]{\singlespace \tiny #1}}
\newcommand{\pois}{Poiseuille}
\newcommand{\Hs}      {\hspace{-0.5cm}} %
\newcommand{\CIF}     {\centering \includegraphics[width=2.7in]} %
\newcommand{\Vmin}    {\vspace{-0.2cm}} %

\renewcommand{\sectionmark}[1]%
      {\markright{\thesection\ #1}} 

\renewcommand{\note}[1]{}

\doublespace 

\title
{ %
\vspace*{3.0cm} \LARGE{\bf Pore-Scale Modeling of Navier-Stokes Flow in Distensible Networks and Porous Media} \vspace*{4.0cm} \\
}

\author{Taha Sochi\footnote{University College London, Department of Physics \& Astronomy, Gower Street, London, WC1E 6BT.
Email: t.sochi@ucl.ac.uk.} \vspace*{5.0cm}}

\begin{document}

\maketitle %
\pagenumbering{arabic}

\newpage
\phantomsection \addcontentsline{toc}{section}{Contents} %
\tableofcontents

%

\newpage
\phantomsection \addcontentsline{toc}{section}{Abstract} \noindent
{\noindent \LARGE \bf Abstract} \vspace{0.5cm}\\
\noindent %

In this paper, a pore-scale network modeling method, based on the flow continuity residual in
conjunction with a Newton-Raphson non-linear iterative solving technique, is proposed and used to
obtain the pressure and flow fields in a network of interconnected distensible ducts representing,
for instance, blood vasculature or deformable porous media. A previously derived analytical
expression correlating boundary pressures to volumetric flow rate in compliant tubes for a
pressure-area constitutive elastic relation has been used to represent the underlying flow model.
Comparison to a preceding equivalent method, the one-dimensional Navier-Stokes finite element, was
made and the results were analyzed. The advantages of the new method have been highlighted and
practical computational issues, related mainly to the rate and speed of convergence, have been
discussed.


\pagestyle{headings} %
\addtolength{\headheight}{+1.6pt}
\lhead[{Chapter \thechapter \thepage}]%
      {{\bfseries\rightmark}}
\rhead[{\bfseries\leftmark}]%
     {{\bfseries\thepage}} 
\headsep = 1.0cm               

\newpage
\section{Introduction}

There are many scientific, industrial and biomedical applications related to the flow of fluids in
distensible networks of interconnected tubes and compliant porous materials. A few examples are
magma migration, microfluidic sensors, fluid filtering devices, deformable porous geological
structures such as those found in petroleum reservoirs and aquifers, as well as almost all the
biological flow phenomena like blood circulation in the arterial and venous vascular trees or
biological porous tissue and air movement in the lung windpipes.

There have been many studies in the past related to this subject \cite{Whitaker1986,
Spiegelman1993, ChenELQSY2000, Ambrosi2002, FormaggiaLQ2003, KrenzD2003, KlubertanzBLV2003,
MabotuwanaCP2007, Nobile2009, JuWF2011}; however most of these studies are based on complex
numerical techniques built on tortuous mathematical infrastructures which are not only difficult to
implement with expensive computational running costs, but are also difficult to verify and
validate. The widespread approach in modeling the flow in deformable structures is to use the
one-dimensional Navier-Stokes finite element formulation for modeling the flow in networks of
compliant large tubes \cite{FormaggiaLQ2003, SochiTechnical1D2013} and the extended Darcy
formulation for the flow in deformable porous media which is based on the poromechanics theory or
some similar numerical meshing techniques \cite{HuygheACR1992, CoussyBook2004, CoussyBook2010,
ChapelleGMC2010}. Rigid network flow models, like \pois, may also be used as an approximation
although in most cases this is not really a good one \cite{SochiPois1DComp2013}.

There have also been many studies in the past related to the flow of fluids in ensembles of
interconnected ducts using pore-scale network modeling especially in the earth science and
petroleum engineering disciplines \cite{SorbieCJ1989, Sorbie19902, Sorbiebook1991, BluntJPV2002,
Valvatnethesis2004, Lopezthesis2004, Sochithesis2007, SochiB2008, SochiVE2009, SochiYield2010}.
However, there is hardly any work on the use of pore-scale network modeling to simulate the flow of
fluids in deformable structures with distensible characteristics such as elastic or viscoelastic
mechanical properties.

There are several major advantages in using pore-scale network modeling over the more traditional
analytical and numerical approaches. These advantages include a comparative ease of implementation,
relatively low computational cost, reliability, robustness, relatively smooth convergence, ease of
verification and validation, and obtaining results which are usually very close to the underlying
analytical model that describes the flow in the individual ducts. Added to all these a fair
representation and realistic description of the flow medium and the essential physics at
macroscopic and mesoscopic levels \cite{SochiFeature2010, SochiComp2010}. Pore-scale modeling, in
fact, is a balanced compromise between the technical complexities and the physical reality. More
details about pore-scale network modeling approach can be found, for instance, in \cite{Blunt2001,
SochiArticle2010}.

In this paper we use a residual-based non-linear solution method in conjunction with an analytical
expression derived recently \cite{SochiElastic2013} for the one-dimensional Navier-Stokes flow in
elastic tubes to obtain the pressure and flow fields in networks of interconnected distensible
ducts. The residual-based scheme is a standard method for solving systems of non-linear equations
and hence is commonly used in fluid mechanics for solving systems of partial differential equations
obtained, for example, in a finite element formulation \cite{SochiTechnical1D2013}. The proposed
method is based on minimizing the residual obtained from the conservation of volumetric flow rate
on the individual network nodes with a Newton-Raphson non-linear iterative solution scheme in
conjunction with the aforementioned analytical expression. Other analytical, empirical and even
numerical relations \cite{SochiVariational2013} describing the flow in deformable ducts can also be
used to characterize the underlying flow model.

\section{Method} \label{Method}

The flow of an incompressible Newtonian fluid in a tube with length $L$ and cross sectional area
$A$ assuming a laminar axi-symmetric slip-free flow with a fixed profile and negligible
gravitational body forces can be described by the following one-dimensional Navier-Stokes system of
mass and momentum conservation relations
\begin{eqnarray}
\frac{\partial A}{\partial t}+\frac{\partial Q}{\partial x}&=&0\,\,\,\,\,\,\,\,\,\,\,\,\,
t\ge0,\,\,\, x\in[0,L]     \\
\frac{\partial Q}{\partial t}+\frac{\partial}{\partial x}\left(\frac{\alpha
Q^{2}}{A}\right)+\frac{A}{\rho}\frac{\partial p}{\partial
x}+\kappa\frac{Q}{A}&=&0\,\,\,\,\,\,\,\,\,\,\,\,\, t\ge0,\,\,\, x\in[0,L]     \label{bbb}
\end{eqnarray}
where $Q$ is the volumetric flow rate, $t$ is the time, $x$ is the axial coordinate along the tube,
$\alpha$ is the momentum flux correction factor, $\rho$ is the fluid mass density, $p$ is the local
pressure, and $\kappa$ is the viscosity friction coefficient which is normally given by $\kappa
=\frac{2\pi\alpha\nu}{\alpha-1}$ with $\nu$ being the fluid kinematic viscosity defined as the
ratio of the dynamic viscosity $\mu$ to the mass density \cite{BarnardHTV1966, SochiSlip2011,
SochiTechnical1D2013, SochiPois1DComp2013, SochiNavier2013}. These relations are usually supported
by a constitutive relation that correlates the pressure to the cross sectional area in a
distensible tube, to close the system in the three variables $A$, $Q$ and $p$.

The usual method for solving this system of equations for a single compliant tube in transient and
steady state flow is to use the finite element method based on the weak formulation by multiplying
the mass and momentum conservation equations by weight functions and integrating over the solution
domain to obtain the weak form of the system. This weak form, with suitable boundary conditions,
can then be used as a basis for finite element implementation in conjunction with an iterative
scheme such as Newton-Raphson method. The finite element system can also be extended to a network
of interconnected deformable tubes by imposing suitable boundary conditions, based on pressure or
flux constraints for instance, on all the boundary nodes, and coupling conditions on all the
internal nodes. The latter conditions are normally derived from Riemann's method of
characteristics, and the conservation principles of mass and mechanical energy in the form of
Bernoulli equation for inviscid flow with negligible gravitational body forces
\cite{SochiBranchFlow2013}. More details on the finite element formulation, validation and
implementation are given in \cite{SochiTechnical1D2013}.

The pore-scale network modeling method, which is proposed as a substitute for the finite element
method in steady state flow, is established on three principles: the continuity of mass represented
by the conservation of volumetric flow rate for incompressible flow, the continuity of pressure
where each branching nodal point has a uniquely defined pressure value \cite{SochiBranchFlow2013},
and the characteristic relation for the flow of the specific fluid model in the particular
structural geometry such as the flow of power law fluids in rigid tubes or the flow of Newtonian
fluids in elastic ducts. The latter principle is essentially a fluid-structure interaction
attribute of the adopted flow model especially in the context of compliant ducts.

In more technical terms, the pore-scale network modeling method employs an iterative scheme for
solving the following matrix equation which is based on the flow continuity residual

\begin{equation}\label{JDpR1}
\mathbf{J}\Delta\mathbf{p}=-\mathbf{r}
\end{equation}
where $\mathbf{J}$ is the Jacobian matrix, $\mathbf{p}$ is the vector of variables which represent
the pressure values at the boundary and branching nodes, and $\mathbf{r}$ is the vector of
residuals which is based on the continuity of the volumetric flow rate. For a network of
interconnected tubes defined by $n$ boundary and branching nodes the above matrix equation is
defined by

\begin{equation}\label{JDpR2}
\left[\begin{array}{ccc}
\frac{\partial f_{1}}{\partial p_{1}} & \cdots & \frac{\partial f_{1}}{\partial p_{n}}\\
\vdots & \ddots & \vdots\\
\frac{\partial f_{n}}{\partial p_{1}} & \cdots & \frac{\partial f_{n}}{\partial
p_{n}}\end{array}\right]\left[\begin{array}{c}
\Delta p_{1}\\
\vdots\\
\Delta p_{n}\end{array}\right]=\left[\begin{array}{c}
r_{1}\\
\vdots\\
r_{n}\end{array}\right]
\end{equation}
where the subscripts stand for the nodal indices, $p$ and $r$ are the nodal pressure and residual
respectively, and $f$ is the flow continuity residual function which, for a general node $j$, is
given by

\begin{equation}
f_{j}=\sum_{i=1}^{m}Q_{i}=0
\end{equation}

In the last equation, $m$ is the number of flow ducts connected to node $j$, and $Q_{i}$ is the
volumetric flow rate in duct $i$ signed ($+/-$) according to its direction with respect to the
node, i.e. toward or away. For the boundary nodes, the continuity residual equations are replaced
by the boundary conditions which are usually based on the pressure or flow rate constraints. In the
computational implementation, the Jacobian is normally evaluated numerically by finite differencing
\cite{SochiTechnical1D2013}.

The procedure to obtain a solution by the residual-based pore-scale modeling method starts by
initializing the pressure vector $\mathbf{p}$ with initial values. Like any other numerical
technique, the rate and speed of convergence is highly dependent on the initial values of the
variable vector. The system given by Equation \ref{JDpR2} is then constructed where the Jacobian
matrix and the residual vector are calculated in each iteration. The system \ref{JDpR2} is then
solved for $\Delta\mathbf{p}$, i.e.

\begin{equation}
\Delta\mathbf{p}=-\mathbf{J}^{-1}\mathbf{r}
\end{equation}
and the vector $\mathbf{p}$ in iteration $l$ is updated to obtain a new pressure vector for the
next iteration ($l+1$), that is

\begin{equation}
\mathbf{p}_{l+1}=\mathbf{p}_{l}+\Delta\mathbf{p}
\end{equation}

This is followed by computing the norm of the residual vector from the following equation

\begin{equation}
\mathfrak{N}=\frac{\sqrt{r_{1}^{2}+\cdots+r_{n}^{2}}}{n}
\end{equation}
where $r$ is the flow continuity residual. This cycle is repeated until the norm is less than a
predefined error tolerance or a certain number of iteration cycles is reached without convergence.
In the last case, the operation will be deemed a failure and hence it will be aborted to be resumed
possibly with improved initial values or even modified model parameters if the physical problem is
flexible and allows for a certain degree of freedom.

The characteristic flow relation that has to be used for computing $Q$ in the residual equation is
dependent on the flow model. As for the flow of Newtonian fluids in distensible tubes based on the
previously-described system of flow equations, the following analytical relation representing the
one-dimensional Navier-Stokes flow in elastic tubes can be used

\begin{equation}\label{QElastic2}
Q=\frac{-\kappa L+\sqrt{\kappa^{2}L^{2}-4\alpha\ln\left(A_{in}/A_{ou}\right)\frac{\beta}{5\rho
A_{o}}\left(A_{ou}^{5/2}-A_{in}^{5/2}\right)}}{2\alpha\ln\left(A_{in}/A_{ou}\right)}
\end{equation}
Other analytical or empirical or numerical relations characterizing the flow rate can also be used
in this context \cite{SochiVariational2013}.

The flow relation of Equation \ref{QElastic2} was previously derived and validated by a
one-dimensional finite element method in \cite{SochiElastic2013}. Equation \ref{QElastic2} is based
on a pressure-area constitutive elastic relation in which the pressure is proportional to the
radius change with a proportionality stiffness factor that is scaled by the reference area, i.e.

\begin{equation}\label{pAEq2}
p=\frac{\beta}{A_{o}}\left(\sqrt{A}-\sqrt{A_{o}}\right)
\end{equation}

In the last two equations, $A_{o}$ is the reference area corresponding to the reference pressure
which in this equation is set to zero for convenience without affecting the generality of the
results, $A_{in}$ and $A_{ou}$ are the tube cross sectional area at the inlet and outlet
respectively, $A$ is the tube cross sectional area at the actual pressure, $p$, as opposed to the
reference pressure, and $\beta$ is the tube wall stiffness coefficient which is usually defined by
\begin{equation}
\beta=\frac{\sqrt{\pi}h_oE}{1-\varsigma^2}
\end{equation}
where $h_o$ is the tube wall thickness at reference pressure, while $E$ and $\varsigma$ are
respectively the Young's elastic modulus and Poisson's ratio of the tube wall.

With regard to the validation of the numeric solutions obtained from the finite element and
pore-scale methods, the time independent solutions of the one-dimensional finite element model can
be tested for validation by satisfying the boundary and coupling conditions as well as the analytic
solution given by Equation \ref{QElastic2} on each individual duct, while the solutions of the
residual-based pore-scale modeling method are validated by testing the boundary conditions and the
continuity of volumetric flow rate at each internal node, as well as the analytic solution given by
Equation \ref{QElastic2} which is inevitably satisfied if the continuity equation is satisfied
according to the pore-scale solution scheme. The necessity to satisfy the analytic solution on each
individual tube in the finite element method is based on the fact that the flow in the individual
tubes according to the underlying one-dimensional model is dependent on the imposed boundary
conditions but not on the mechanism by which these conditions are imposed. In the case of finite
element with tube discretization and/or employing non-linear interpolation orders, the solution at
the internal points of the ducts can also be tested by satisfying the following analytical relation
\cite{SochiTechnical1D2013}

\begin{equation}\label{AnalEq}
x=\frac{-\alpha Q^{2}\ln(A/A_{\mathrm{in}})+\beta\left(A^{5/2}-A_{\mathrm{in}}^{5/2}\right)/(5\rho
A_{o})}{-\left[2\pi\alpha\nu/(\alpha-1)\right]Q}
\end{equation}
The derivation of this equation is similar to the derivation of Equation \ref{QElastic2} but with
using the inlet boundary condition only. In fact even Equation \ref{QElastic2} can be used for
testing the solution at the internal points if we assume these points as periphery nodes
\cite{SochiTechnical1D2013}.

\section{Implementation and Results}

The residual-based pore-scale modeling method, as described in the last section, was implemented in
a computer code with an iterative Newton-Raphson method that includes four numeric solvers (SPARSE,
SUPERLU, UMFPACK, and LAPACK). The code was then tested on computer-generated networks representing
distensible fluid transportation structures like ensembles of interconnected tubes or porous media.
A sample of these networks are given in Figure \ref{Networks}. Because the residual-based
pore-scale method can be used in general to obtain flow solutions for any characteristic flow that
involves linear or non-linear fluid models, such as Newtonian or non-Newtonian fluids, passing
through rigid or distensible networks, the code was tested first on \pois\ and power law fluids in
rigid networks \cite{SochiPower2011, SochiPois1DComp2013}. The results for these validation tests
were exceptionally accurate with very low error margin over the whole network and with smooth
convergence.

We also used the one-dimensional finite element model that we briefly described in the last section
for the purpose of comparison. This model was previously implemented in a computer code with a
residual-based Newton-Raphson iterative solution scheme, similar to the one used in the pore-scale
modeling. Full description of the finite element method, code and techniques can be found in
\cite{SochiTechnical1D2013}. A number of pore-scale and one-dimensional finite element time
independent flow simulations were carried out on our computer-generated networks using a range of
physical parameters defining the fluid and structure as well as different numeric solvers with
different solving schemes. Various types of pressure and flow boundary conditions were imposed in
these simulations, although in most cases Dirichlet-type pressure boundary conditions were applied.
The finite element simulations were performed using a linear Lagrange interpolation scheme with no
tube discretization to closely match the pore-scale modeling approach. All the results of the
reported and unreported runs have passed the rigorous validation tests that we stated earlier in
section \ref{Method}.

Regarding the nature of the networks, several types of networks have been generated and used in the
above-mentioned flow simulations; these include fractal, cubic and orthorhombic networks. The
fractal networks are based on fractal branching patterns where each generation of the branching
tubes in the network have a specific number of branches related to the number of branches in the
parent generation, such as 2:1, as well as specific branching angle, radius branching ratio and
length to radius ratio. The radius branching ratio is normally based on a Murray-type relation
\cite{Murray1926a, Murray1926b, Murray1926c, SochiBranchFlow2013}. The fractal networks are also
characterized by the number of generations. The fractal networks used in this study have a single
inlet boundary node and multiple outlet boundary nodes \cite{SochiPois1DComp2013}.

The cubic and orthorhombic networks are based on a cubic or orthorhombic three-dimensional lattice
structure where the radii of the tubes in the network can be constant or subject to statistical
random distributions such as uniform distribution. These networks have a number of inlet boundary
nodes on one side and a similar number of outlet boundary nodes on the opposite side while the
nodes on the other four sides (i.e. the lateral) are considered internal nodes. The boundary
conditions are then imposed on these inlet and outlet boundary nodes individually according to the
need. A sample of the fractal, cubic and orthorhombic networks used in this investigation are shown
in Figure \ref{Networks}.

It is noteworthy that all the networks used in the pore-scale and finite element flow simulations
consist of interconnected straight cylindrical tubes where each tube is characterized by a constant
radius over its entire length and spatially identified by two end nodes. These networks are totally
connected, that is any node in the network can be reached from any other node by moving entirely
inside the network space. As for the physical size, we used different sizes to represent different
flow structures such as arterial and venous blood vascular trees and spongy porous geological
media. The physical parameters used in our simulations, especially those related to the fluids and
flow structures, were generally selected to represent realistic physical systems although physical
parameters representing hypothetical conditions have also been used for the purpose of test and
validation. However, since the current study is purely theoretical with no involvement of
experimental or observational data, the validity of the reported models are not affected by the
actual values of the physical parameters although this has some consequences on the speed and rate
of convergence in different physical regimes.

\begin{figure}[!h]
\centering{}
\includegraphics[scale=0.75]{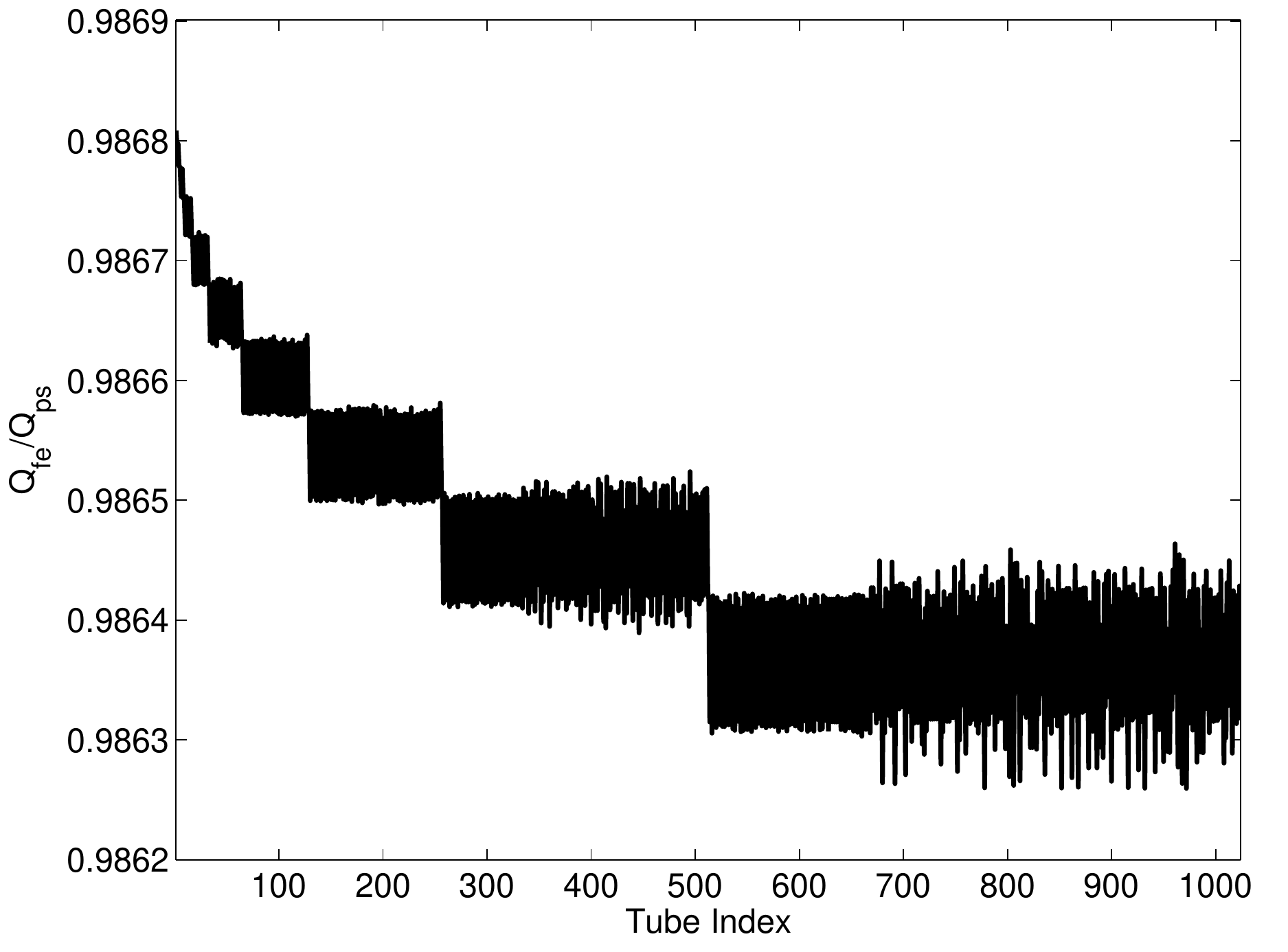}
\caption{The ratio of flow rate of finite element to pore-scale models versus tube index for a
fractal network. The network consists of 10 generations with an area-preserving branching index of
2 \cite{SochiBranchFlow2013} and an inlet main branch with $R=5$~mm and $L=50$~mm. The parameters
for these simulations are: $\beta=236.3$~Pa.m, $\rho=1060.0$~kg.m$^{-3}$, $\mu=0.0035$~Pa.s, and
$\alpha=1.333$. The inlet and outlet pressures are: $p_{i}=2000$~Pa and $p_{o}=0.0$~Pa. The fluid
and structure parameters are chosen to roughly resemble blood circulation in large vessels.}
\label{FractalCompareQ}
\end{figure}

\begin{figure}[!h]
\centering{}
\includegraphics[scale=0.75]{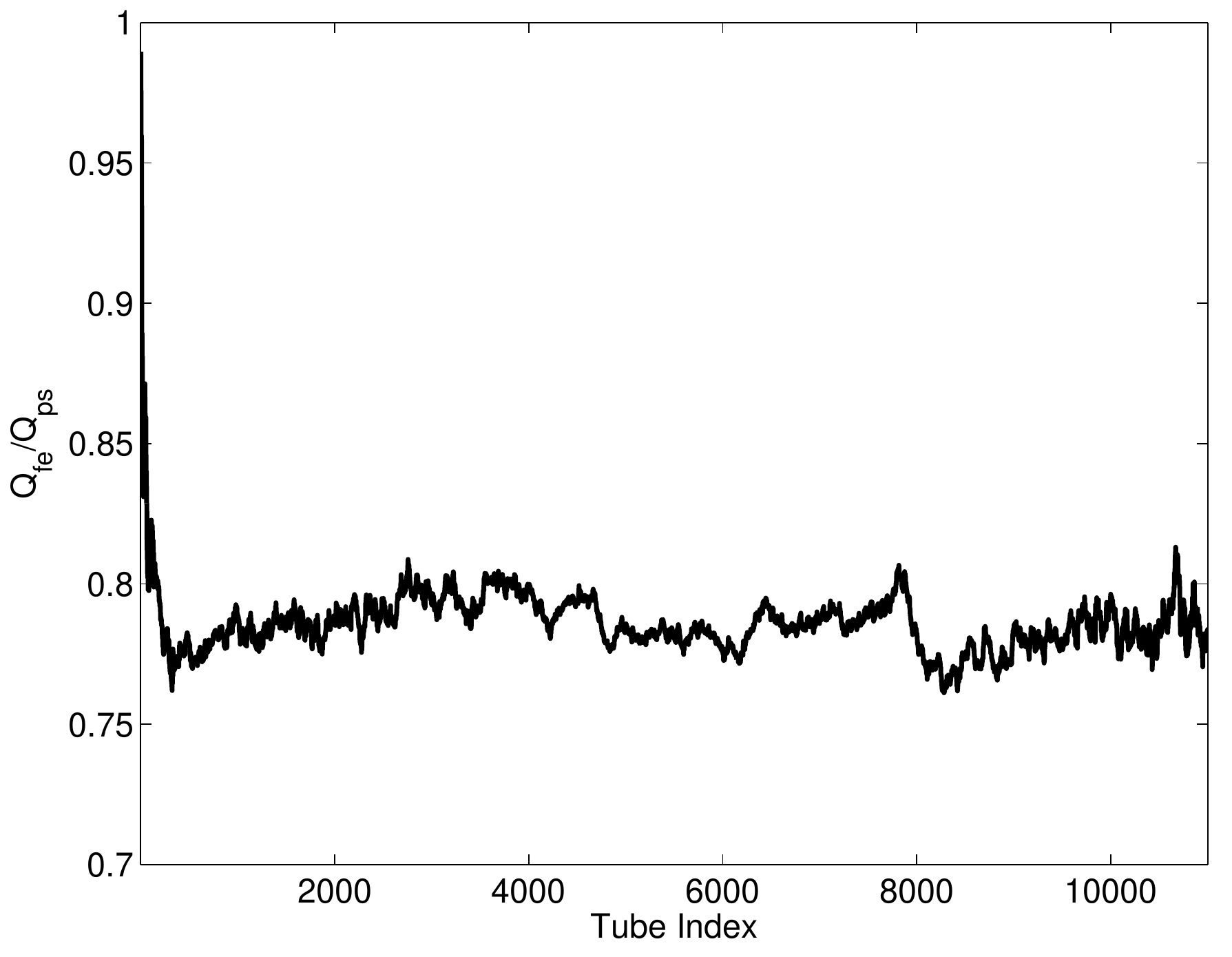}
\caption{The ratio of flow rate of finite element to pore-scale models versus tube index for an
inhomogeneous orthorhombic network. The network consists of about 11000 tubes with different
lengths and length to radius ratios as explained in the main text. The parameters for these
simulations are: $\beta=236.3$~Pa.m, $\rho=860.0$~kg.m$^{-3}$, $\mu=0.075$~Pa.s, and $\alpha=1.2$.
The inlet and outlet pressures are: $p_{i}=3000$~Pa and $p_{o}=0.0$~Pa. The fluid and structure
parameters are chosen to approximately match the transport of crude oil through an elastic
structure.} \label{InhomoRectCompareQ}
\end{figure}

In Figures \ref{FractalCompareQ} and \ref{InhomoRectCompareQ} we present a sample of the above
mentioned comparative simulations. In Figure \ref{FractalCompareQ} we plot the ratio of the flow
rate obtained from the finite element model to that obtained from the pore-scale model for a
fractal network with an area-preserving branching index of 2 \cite{SochiBranchFlow2013} where the
number of tubes in each generation is twice the number in the parent generation. In these flow
simulations we applied an inlet boundary pressure of 2000~Pa on the single inlet boundary node of
the main branch and an outlet boundary pressure of 0.0~Pa on all the outlet boundary nodes. The
results of the pore-scale and finite element models are very close although the two models differ
due to the use of different branching coupling conditions, i.e. continuity of pressure for the
pore-scale model and Bernoulli for the finite element. The effect of the coupling conditions on the
different generations of the fractal network can be seen in Figure \ref{FractalCompareQ} where a
generation-based configuration is obvious. This feature is a clear indication of the effect of the
coupling conditions on the deviation between the two models.

In Figure \ref{InhomoRectCompareQ} we compare the pore-scale and finite element models for an
inhomogeneous orthorhombic network consisting of about 11000 interconnected tubes, similar to the
one depicted in Figure \ref{Networks} (f), where we plotted the ratio of the flow rate obtained
from the two models as for the fractal network. The network is generated with a random uniform
distribution for the tubes radii with a variable length in different orientations and with a
variable length to radius ratio that ranges between about 5-15. The dimensions of the flow
structure are $2\times1.5\times1$~m with a constant inlet boundary pressure of 3000~Pa applied to
all the nodes on the inlet side and zero outlet boundary pressure applied to all the nodes on the
outlet side. The fluid and structure physical parameters for the flow model are selected to roughly
resemble the flow of crude oil in some elastic structure, possibly in a refinement processing
plant. As seen, the pore-scale and finite element results differ significantly on most part of the
network. The reason in our judgment is the effect of the branching coupling conditions (i.e.
continuity of pressure and Bernoulli) which have a stronger impact in such a network than a fractal
network due to the inhomogeneity with random radius distribution on one hand and the high nodal
connectivity of the orthorhombic lattice on the other hand. The fluid property which significantly
differs from that of the fractal network simulation may also have a role in exacerbating the
discrepancy.

The results shown in these figures represent a sample of our simulations which reflect the general
trend in other simulations. However the agreement between the pore-scale and finite element models
is highly dependent on the flow regime and the nature of the physical problem which combines the
fluid, structure and their interaction. As indicated early, the discrepancy between the two models
reflects the effect of the coupling conditions, i.e. the pressure continuity for the pore-scale
model and the Bernoulli condition for the finite element. The gravity of this effect is strongly
dependent on the type of fluid, flow regime, inhomogeneity and connectivity.

It should be remarked that in these figures (i.e. \ref{FractalCompareQ} and
\ref{InhomoRectCompareQ}) we used the volumetric flow rate, rather than the nodal pressure, to make
the comparison. The reason is that comparing the pressure is not possible because nodal pressure is
not defined in the finite element model due to the use of the Bernoulli equation
\cite{SochiBranchFlow2013} where each node has a number of pressure values matching the number of
the connected tubes.


\begin{figure}
\centering %
\subfigure[11-generation fractal]%
{\begin{minipage}[b]{0.5\textwidth} \CIF {Fractal11}
\end{minipage}}
\Hs %
\subfigure[13-generation fractal]%
{\begin{minipage}[b]{0.5\textwidth} \CIF {Fractal13}
\end{minipage}} \Vmin

%
\centering %
\subfigure[homogeneous cubic]%
{\begin{minipage}[b]{0.5\textwidth} \CIF {RegularSquare}
\end{minipage}}
\Hs %
\subfigure[inhomogeneous cubic]%
{\begin{minipage}[b]{0.5\textwidth} \CIF {IrregularSquare}
\end{minipage}} \Vmin

%
\centering %
\subfigure[homogeneous orthorhombic]%
{\begin{minipage}[b]{0.5\textwidth} \CIF {RegularRectangular}
\end{minipage}}
\Hs %
\centering %
\subfigure[inhomogeneous orthorhombic]%
{\begin{minipage}[b]{0.5\textwidth} \CIF {IrregularRectangular}
\end{minipage}}
\caption{A sample of computer-generated fractal, cubic and orthorhombic networks used in the
current investigation. \label{Networks}}
\end{figure}


\section{Pore-Scale vs. Finite Element}

It is difficult to make an entirely fair comparison between the pore-scale and finite element
methods due mainly to the use of different coupling conditions at the branching junctions as well
as different theoretical assumptions. Therefore, the pressure and flow rate fields obtained from
these two methods on a given network are generally different. The difference, however, is highly
dependent on the nature of the specified physical and computational conditions.

One of the advantages of the pore-scale modeling method over the finite element method, in addition
to the general advantages of the pore-scale modeling approach which were outlined earlier, is that
when pore-scale method converges it usually converges to the underlying analytic solution with
negligible marginal errors over the whole network, while the finite element method normally
converges with significant errors over some of the network ducts especially those with eccentric
geometric characteristics such as very low length to radius ratio \cite{SochiTechnical1D2013}. It
may also be argued that the coupling condition used in the pore-scale modeling method, which is
based on the continuity of pressure, is better than the corresponding coupling condition used in
the finite element method which is based on the Bernoulli inviscid flow with discontinuous pressure
at the nodal points. Some of the criticism to the use of Bernoulli as a coupling condition is
outlined in \cite{SochiBranchFlow2013}. Another advantage of the pore-scale modeling is that it is
generally more stable than the finite element with a better convergence behavior due partly to the
simpler pore-scale computational infrastructure.

The main advantage of the finite element method over the pore-scale modeling method is that it
accommodates time dependent flow naturally, as well as time independent flow, while pore-scale
modeling in its current formulation is capable only of dealing with time independent flow.
Moreover, the finite element method may be better suited for describing other one-dimensional
transportation phenomena such as wave propagation and reflection in deformable networks. However,
time dependent flow can be simulated within the pore-scale modeling framework as a series of time
independent frames although this is not really a time dependent flow but rather a pseudo time
dependent. Another advantage of the finite element method is that it is capable, through the use of
segment discretization and higher orders of interpolation, of computing the pressure and flow rate
fields at the internal points along the tubes length and not only on the tubes periphery points at
the nodal junctions. However, due to the incompressibility of the flow, computing the flow rate at
the internal points is redundant as it is identical to the flow rate at the end points. With regard
to computing the pressure field on the internal points, it can also be obtained by pore-scale
modeling method through the application of Equation \ref{AnalEq} to the solution obtained on the
individual ducts. Moreover, it can be obtained by creating internal nodal junctions along the tubes
through the use of tube discretization, similar to the discretization in the finite element method.

With regard to the size of the problem, which directly influences the ensuing memory cost as well
as the CPU time, the number of degrees of freedom for the pore-scale model is half the number of
degrees of freedom for the one-dimensional finite element model due to the fact that the former has
one variable only ($p$) while the latter has two variables ($p$ and $Q$). This estimation of the
finite element computational cost is based on using a linear interpolation scheme with no tube
discretization; and hence this cost will substantially increase with the use of discretization
and/or higher orders of interpolation. The computational cost for both models also depends on the
type of the solver used such as being sparse or dense, and direct or iterative, as well as some
problem-specific implementation overheads.

CPU processing time depends on several factors such as the size and type of the network, the
initial values for the flow solutions, the parameters of the fluid and tubes, the employed
numerical solver, and the assumed pressure-area constitutive relation. Typical processing time for
a single run of the pore-scale network model on a typical laptop or desktop computer ranges between
a few seconds to few minutes using a single processor with an average speed of 2-3 gigahertz. The
CPU processing time for the time independent finite element model is comparable to the processing
time of the pore-scale network model. In both cases, the final convergence in a typical problem is
normally reached within 3-7 Newton-Raphson iterations depending mainly on the initial values.

\section{Convergence Issues}

Like the one-dimensional Navier-Stokes finite element model, the residual-based pore-scale method
may suffer from convergence difficulties due to the highly non-linear nature of the flow model. The
nonlinearity increases, and hence the convergence difficulties aggravate, with increasing the
pressure gradient across the flow domain. The nonlinearity also increases with eccentric values
representing the fluid and structure parameters such as the fluid viscosity or wall distensibility.
Several numerical tricks and stabilization techniques can be used to improve the rate and speed of
convergence. These include non-dimensionalization of the flow equations, using a variety of unit
systems such as m.kg.s or mm.g.s or m.g.s for the input data and parameters, and scaling the
network flow model up or down to obtain a similarity solution that can be scaled back to obtain the
final solution. The error tolerance for the convergence criterion which is based on the residual
norm may also be increased to enhance the rate and speed of convergence. Despite the fact that the
use of relatively large error tolerance can cause a convergence to a wrong solution or to a
solution with large errors, the solution can always be tested by the above-mentioned validation
metrics and hence it is accepted or rejected according to the adopted approval criteria
\cite{SochiTechnical1D2013}.

Other convergence-enhancing methods can also be used. In the highly non-linear cases, the initial
values to initiate the variable vector can be obtained from a \pois\ solution which can be easily
acquired within the same code. The convergence, as indicated already, becomes more difficult with
increasing the pressure gradient across the flow domain, due to an increase in the nonlinearity. An
effective approach to obtain a solution in such cases is to step up through a pressure ladder by
gradual increase in the pressure gradient where the solution obtained from one step is used as an
initial value for the next step. Although this usually increases the computational cost, the
increase in most cases is not substantial because the convergence becomes rapid with the use of
good initial values that are close to the solution. The convergence rate and speed may also be
improved by adjusting the flow parameters. Although the parameters are dependent on the nature of
the physical problem and hence they are not a matter of choice, there may be some freedom in tuning
some non-critical parameters. In particular, adjusting the correction factor for the axial momentum
flux, $\alpha$, can improve the convergence and quality of solution. The rate and speed of
convergence may also depend on the employed numerical solver.

Another possible convergence trick is to use a large error margin for the residual norm to obtain
an approximate solution which can be used as an initial guess for a second run with a smaller error
margin. On repeating this process, with progressively reducing the error margin, a reasonably
accurate solution can be obtained eventually. It should be remarked that the pore-scale and finite
element models have generally different convergence behaviors where each converges better than the
other for certain flow regimes or fluid-structure physical problems. However, in general the
pore-scale model has a better convergence behavior with a smaller error, as indicated earlier.
These issues, however, are strongly dependent on the implementation and practical coding aspects.

\section{Conclusions}

In this paper, a pore-scale modeling method has been proposed and used to obtain the pressure and
flow fields in distensible networks and porous media. This method is based on a residual
formulation obtained from the continuity of volumetric flow rate at the branching junctions with a
Newton-Raphson iterative numeric technique for solving a system of simultaneous non-linear
equations. An analytical relation linking the flow rate in distensible tubes to the boundary
pressures is exploited in this formulation. This flow relation is based on a pressure-area
constitutive equation derived from elastic tube deformability characteristics.

A comparison between the proposed pore-scale network modeling approach and the traditional
one-dimensional Navier-Stokes finite element approach has also been conducted with a main
conclusion that pore-scale modeling method has obvious practical and theoretical advantages,
although it suffers from some limitations related mainly to its static time independent nature. We
therefore believe that although the pore-scale modeling approach cannot totally replace the
traditional methods for obtaining the flow rate and pressure fields over networks of interconnected
deformable ducts, it is a valuable addition to the tools used in such flow simulation studies.

\clearpage
\phantomsection \addcontentsline{toc}{section}{Nomenclature} %
{\noindent \LARGE \bf Nomenclature} \vspace{0.5cm}

\begin{supertabular}{ll}
$\alpha$                &   correction factor for axial momentum flux \\
$\beta$                 &   stiffness coefficient in pressure-area relation \\
$\kappa$                &   viscosity friction coefficient \\
$\mu$                   &   fluid dynamic viscosity \\
$\nu$                   &   fluid kinematic viscosity \\
$\rho$                  &   fluid mass density \\
$\varsigma$             &   Poisson's ratio of tube wall \\
\\
$A$                     &   tube cross sectional area \\
$A_{in}$                &   tube cross sectional area at inlet \\
$A_o$                   &   tube cross sectional area at reference pressure \\
$A_{ou}$                &   tube cross sectional area at outlet \\
$E$                     &   Young's elastic modulus of tube wall \\
$f$                     &   flow continuity residual function \\
$h_o$                   &   vessel wall thickness at reference pressure \\
$\mathbf{J}$            &   Jacobian matrix \\
$L$                     &   tube length \\
$n$                     &   number of network nodes \\
$\mathfrak{N}$          &   norm of residual vector \\
$p$                     &   pressure \\
$\mathbf{p}$            &  pressure vector \\
$p_i$                     &  inlet pressure \\
$p_o$                     &  outlet pressure \\
$\Delta\mathbf{p}$      &   pressure perturbation vector \\
$Q$                     &   volumetric flow rate \\
$r$                     &   flow continuity residual \\
$\mathbf{r}$             &  residual vector \\
$R$                     &   tube radius \\
$t$                     &   time \\
$x$                     &   tube axial coordinate \\

\end{supertabular}

\clearpage
\phantomsection \addcontentsline{toc}{section}{References} %
\bibliographystyle{unsrt}

\end{document}